\documentclass{cta-author}

\usepackage{url}
\usepackage{array}
\usepackage{booktabs}
\usepackage{multirow}
\usepackage{fancyhdr}

{}
{}
{}

\begin{document}

\title{Automated Brain Tumour Segmentation Using Deep Fully Residual Convolutional Neural Networks}

\author{\au{Indrajit Mazumdar$^{1\corr}$}}

\address{\add{1}{Department of Computer Science and Engineering, Indian Institute of Technology Kharagpur, Kharagpur, India}
\email{indrajit\_mazumdar@iitkgp.ac.in}}

\begin{abstract}
Automated brain tumour segmentation has the potential of making a massive improvement in disease diagnosis, surgery, monitoring and surveillance. However, this task is extremely challenging. Here, we describe our automated segmentation method using 2D CNNs that are based on U-Net. To deal with class imbalance effectively, we have used a weighted Dice loss function. We found that increasing the depth of the `{U}' shape beyond a certain level results in a decrease in performance, so it is essential to choose an optimum depth.  We also found that 3D contextual information cannot be captured by a single 2D network that is trained with patches extracted from multiple views whereas an ensemble of three 2D networks trained in multiple views can effectively capture the information and deliver much better performance. We obtained Dice scores of 0.79 for enhancing tumour, 0.90 for whole tumour, and 0.82 for tumour core on the BraTS 2018 validation set. Our method using 2D network consumes very less time and memory, and is much simpler and easier to implement compared to the state-of-the-art methods that used 3D networks; still, it manages to achieve comparable performance to those methods.
\end{abstract}

\maketitle

\thispagestyle{fancy}
\pagestyle{fancy}
\fancyhf{}
\fancyhead[L]{\ifthenelse{\value{page}=1}{}{}}
\fancyfoot[LE,RO]{\thepage}
\renewcommand{\headrulewidth}{0pt}
\renewcommand{\footrulewidth}{0pt}


\section{Introduction}\label{sec1}

The tumours present in the brain are one of the major lethal types of tumour \cite{DeAngelis2001}. The most frequently occurring primary brain tumours are gliomas \cite{Ohgaki2005, Goodenberger2012}. They originate from the neuroglia within the brain \cite{Goodenberger2012,Bauer2013}. A considerable amount of research has been done in this area, but there remains a great need for improving patient treatment. There are mainly two categories of gliomas, namely high-grade gliomas (HGG) and low-grade gliomas (LGG) \cite{Louis2016, Bauer2013}. HGGs are malignant and more aggressive. They have a worse prognosis and patients having HGG survives on average two years or lower \cite{Ohgaki2005}. LGGs are generally benign but may get converted into HGG. They are less aggressive and have a better prognosis. Patients having LGG survives on average for several years \cite{Ohgaki2005}. Gliomas have four types of histologically heterogeneous tumoural structures, namely peritumoral oedema (ED), enhancing core (ET), necrotic core (NCR), and non-enhancing core (NET) \cite{Menze2015}. These tumoural structures are organised into three mutually inclusive tumour sub-regions, namely, enhancing tumour (ET), tumour core (TC), and whole tumour (WT) \cite{Menze2015}. The TC contains within it the ET, NCR, and NET. The WT contains within it the TC and ED. Among the various imaging modalities available, MRI is preferred for imaging of the brain \cite{Isin2016}. Multiple MRI modalities are used to emphasise the different sub-regions of glioma. The different MRI modalities used are T2-weighted (T2), Fluid Attenuated Inversion Recovery (FLAIR), T1-weighted (T1), and contrast-enhanced T1-weighted (T1ce) \cite{Menze2015}. T1ce emphasises the enhancing tumour. T1 and T1ce emphasise the tumour core. T2 and FLAIR emphasise the whole tumour.

Automated segmentation of brain tumours is essential for quantitative assessment of brain tumours. It is much more accurate and objective compared to qualitative assessment, which is subjective. Thus, automated segmentation has the potential of making a massive improvement in diagnosis, surgery, treatment planning, and monitoring the disease. It is much faster than manual segmentation, which makes it scalable and ensures that patients receive more rapid treatment. Automated segmentation systems are objective as they are not subject to intra-rater and inter-rater variations. Because automated segmentation does not depend on radiologists, it results in reducing human labour, which in turn reduces cost. Automated brain tumour segmentation is highly clinically-relevant since accurate segmentation of tumour is required for extracting accurate radiomic features that are used to predict the patient's overall survival  \cite{Bakas2018}.

Automated brain tumour segmentation is extremely difficult and challenging. Brain tumours are highly heterogeneous, having a significant variation in shape, size, and location among patients, which makes prior information about these things not valuable for segmentation. Multiple tumours of different grades may reside simultaneously inside the brain. The boundary between tumour and adjacent healthy tissue is usually unclear since the intensity at the boundaries changes very smoothly. Bias field artefacts and partial volume artefacts are also present in the MRI images. The problem of brain tumour segmentation also suffers from extreme class imbalance.

Generative and discriminative approaches are the two most used approaches for automated segmentation of brain tumours. Generative methods explicitly define the probabilistic distributions of brain tumours \cite{Menze2010}. However, it is extremely challenging and time-consuming to construct an accurate probabilistic model from meaningful analysis of an image. Discriminative methods straightway comprehend the connection between image intensities and labels. In the past, discriminative methods used for brain tumour segmentation employed feature extraction from images followed by classification \cite{Isin2016}. However, it is extremely challenging to select highly representative features for the classifier.

At present, the latest techniques used for automated brain tumour segmentation are discriminative methods that are deep learning-based \cite{Bakas2018}. Such methods based on deep learning can instinctively learn highly representative features and are thus better than traditional discriminative methods. The success achieved by the numerous deep learning networks varies mainly due to the difference in their network architecture and training procedure. Pereira et al. \cite{Pereira2016} used two different 2D CNN architectures;  one for segmenting HGG and the other for segmenting LGG. But their trained CNN predicts the label of the central voxel only within a patch which leads to high memory usage and time consumption. DeepMedic \cite{Kamnitsas2017a} is a 3D CNN that uses two pathways of different resolutions to combine semantic information at distinct scales. Ensembles of Multiple Models and Architectures (EMMA) \cite{Kamnitsas2018}  is an ensemble of widely varying CNN models that include two variations of DeepMedic model \cite{Kamnitsas2017a, Kamnitsas2015}, three variations of 3D FCN model \cite{Long2015}, and two variations of 3D U-Net model \cite{Ronneberger2015}. This network demonstrated good generalisation performance. Wang et al. \cite{Wang2018d} used a cascade approach in which three networks were used to sequentially segment the three different tumour sub-regions wherein each network's output is cropped and fed as input to the next network. They also incorporated 3D contextual information by training their network in three different orthogonal views and averaging the outputs of those networks. However, the cascade approach consumes more time and memory since it is not end-to-end. Isensee et al. \cite{Isensee2018} used a U-Net \cite{Ronneberger2015} inspired 3D CNN. In their network, each encoder block is a residual block \cite{He2016a}. Deep supervision had been used in the decoder part to improve the gradient flow. Myronenko \cite{Myronenko2019} used an asymmetric encoder-decoder network. Notably, they used a variational autoencoder for regularising the encoder. Isensee et al. \cite{Isensee2019} used a modification of the 3D U-Net \cite{Cicek2016} in which they reduced the number of activation maps before each upsampling operation to decrease the memory usage. McKinley et al. \cite{McKinley2019} used U-Net style network in which DenseNet \cite{Huang2017} structures were placed. They also introduced the label-uncertainty loss that can model label noise and uncertainty. Zhou et al. \cite{Zhou2019} used a model cascade approach to segment the three tumour sub-regions sequentially.

\begin{figure*}[b]
\centering{\includegraphics[scale=1.0]{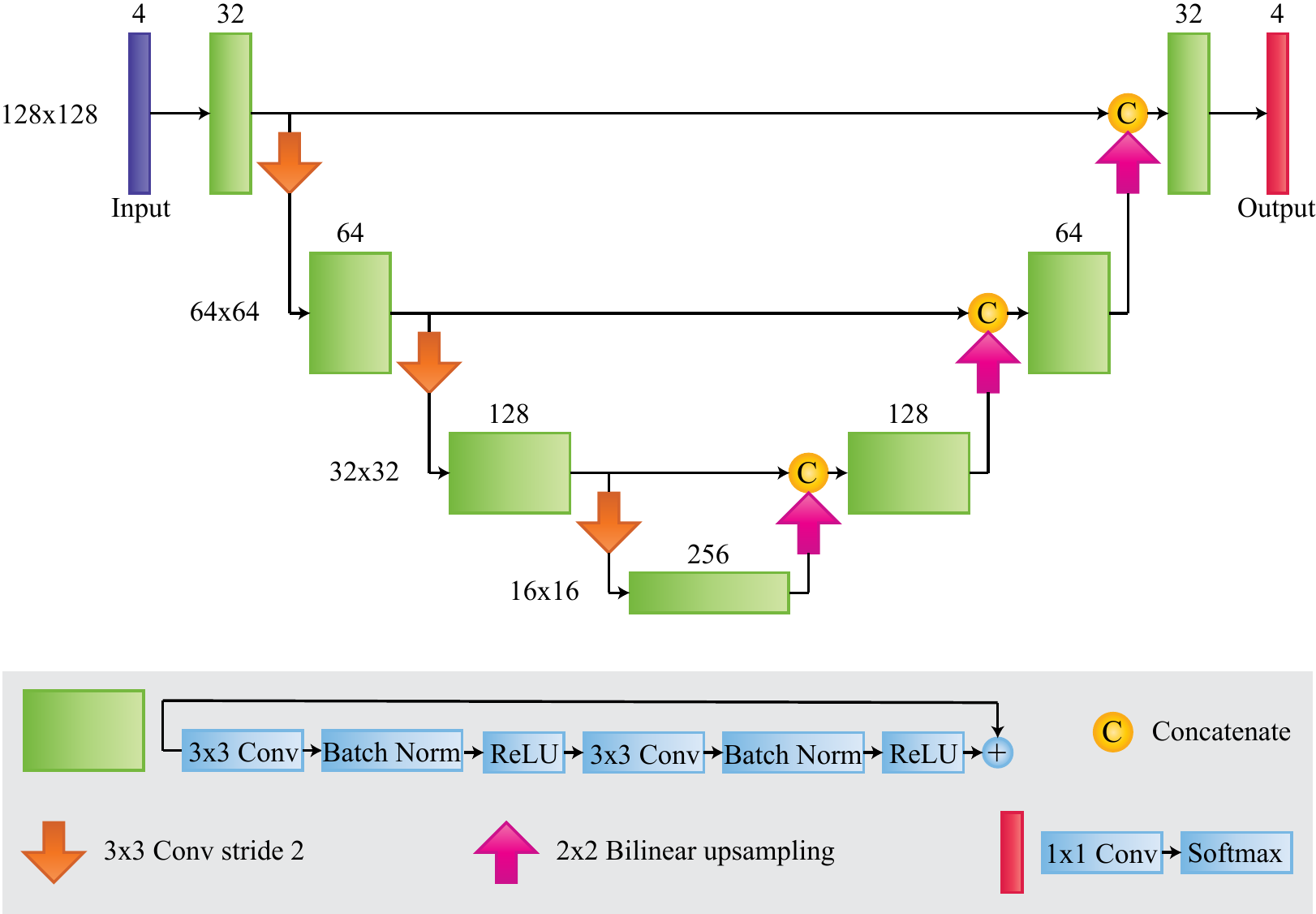}}
\caption{The network architecture illustrated schematically. The violet box denotes input. Each green box corresponds to a residual block. On the top of each box the number channels are denoted. The dimensions of the activation maps is denoted on the left side of each level. Each orange arrow pointing down represents downsampling operation and each pink arrow pointing up represents upsampling operation. Each yellow circle containing $C$ denotes the concatenation of the upsampled activations with the activations from the equivalent level of the encoder. The red box present at the terminal of the decoder represents a $1\times 1$ convolution that is succeeded by a softmax output layer.}
\label{fig1}
\end{figure*}

In this work, we describe our automated glioma segmentation method using 2D CNNs that are based on U-Net \cite{Ronneberger2015}. We have decided to use 2D networks instead of 3D networks since 2D networks have a lower time and memory consumption, and are much simpler and easier to implement compared to 3D networks. We also examine how the performance of a U-Net based network gets affected when we increase the depth of the `{U}' shape and how important it is to choose an optimum depth. We used a weighted Dice loss function to take care of the class imbalance problem effectively. We explore the effectiveness of two different approaches for incorporating 3D contextual information in a 2D network to enhance its performance. In the first approach, we use a single model that is trained by extracting patches from multiple views and in the second approach, we use an ensemble of three networks trained in multiple views. We investigate whether a 2D network that has incorporated 3D contextual information from three orthogonal views can obtain similar performance to the latest 3D networks in the problem of automated segmentation of brain tumours.

We have arranged the remaining portion of this paper in the following way. We describe our proposed segmentation technique in Section~\ref{sec2}. Then, in Section~\ref{sec3}, we deal with the results of the experiments. Eventually, in Section~\ref{sec4}, we conclude this paper.

\section{Methods}\label{sec2}

\subsection{Preprocessing}\label{subsec2.1}

It is essential to normalise the MRI images, so the intensity range remains comparable not only among MRI images of different patients but also among different MRI modalities of the same patient. So, we normalised every MRI modality separately by subtracting the mean and dividing by the standard deviation of the nonzero voxels present inside the region of the brain. The voxels present outside the region of the brain contain only zero intensity.

After normalisation, we extracted 2D slices of size $128\times 128$ from the 3D MRI images. We discarded the patches that do not contain any tumour. Then we shuffled the entire set of extracted patches. After shuffling, we partitioned the set of extracted patches into a training set (80\%) for training our network and an internal validation set (20\%) for parameter tuning.

\subsection{Network Architecture}\label{subsec2.2}

We have designed a U-Net \cite{Ronneberger2015} inspired 2D fully convolutional neural network. We illustrate our network architecture in Fig.~\ref{fig1}. The network is made up of an encoder part to capture context and a symmetric decoder part to enable precise localisation. 

The encoder part is composed of encoder blocks. Each encoder block is a residual block \cite{He2016a} containing two $3\times 3$ convolutions, every one of which is succeeded by a Batch Norm \cite{Ioffe2015} and a ReLU activation. We decreased the resolution of the activation maps by 2 while downsampling by using $3\times 3$ strided convolutions with stride 2. Additionally, we doubled the number of features in every downsampling operation. At the beginning of the encoder part, a total of 32 filters were present in the first convolutional layer. The encoder endpoint has dimensions $16\times 16\times 256$. 

In the decoder part, we upsampled the activation maps using $2\times 2$ bilinear upsampling to increase the resolution of the feature maps by 2. Additionally, we halved the number of features in every upsampling operation. The decoder part is composed of decoder blocks. We have used skip connections for concatenating the upsampled activation maps with the activation maps from the corresponding level of the encoder. Each decoder block has the same structure as that of an encoder block. A $1\times 1$ convolution is used at the end of the decoder to map the feature vector to the four classes, and a softmax output layer follows it.

\begin{table*}[b]
\centering
\caption{Quantitative results on the training set of BraTS 2018.}
\label{tab1}
\renewcommand{\arraystretch}{1.3} 
\begin{tabular}{m{5cm}<{\raggedright\arraybackslash}m{1.5cm}<{\centering}m{1.5cm}<{\centering}m{1.5cm}<{\centering}m{1.5cm}<{\centering}m{1.5cm}<{\centering}m{1.5cm}<{\centering}}
\toprule
\multicolumn{1}{c}{\multirow{2}{*}{Method}} & \multicolumn{3}{c}{Dice score} & \multicolumn{3}{c}{Hausdorff distance (95\% quantile)} \\ \cmidrule(lr){2-4} \cmidrule(l){5-7} 
\multicolumn{1}{c}{} & ET & WT & TC & ET & WT & TC \\ \midrule
Baseline & 0.78 & 0.93 & 0.91 & 5.71 & 11.94 & 9.71 \\
Baseline (depth of `{U}' shape is 4) & 0.74 & 0.90 & 0.89 & 12.12 & 28.23 & 24.52 \\
Baseline (training patches extracted from  three orthogonal views) & 0.80 & 0.94 & 0.91 & 4.96 & 8.31 & 7.29 \\
Baseline (sagittal view) & 0.81 & 0.93 & 0.92 & 5.06 & 13.87 & 5.24 \\
Baseline (coronal view) & 0.78 & 0.93 & 0.92 & 4.36 & 11.82 & 9.51 \\
Ensemble of three models & 0.82 & 0.94 & 0.93 & 2.30 & 5.02 & 2.70 \\ \bottomrule
\end{tabular}
\end{table*}

\subsection{Training}\label{subsec2.3}

Keras \cite{Keras} and Tensorflow \cite{Abadi2016} was used to implement our network. We employed an NVIDIA GeForce GTX TITAN X 12 GB GPU for training our network. The dimensions of training patches were $128\times 128\times 4$, and the batch size was 8. Training continued for 300 epochs. We used Adam optimizer \cite{Kingma2015} and set the initial learning rate to ${{10}^{-4}}$. We applied L2 regularisation with a regularisation strength of ${{10}^{-5}}$ on the convolutional filter weights.

\subsection{Loss Function}\label{subsec2.4}

In the problem of brain tumour segmentation, an extremely small part of the entire image contains tumour. In the training set of the BraTS 2018 \cite{Menze2015, Bakas2017, Bakas2017a, Bakas2017b, Bakas2018} dataset, approximately 98.88\% of total voxels belong to background, 0.64\% belongs to ED, 0.28\% belongs to NCR and NET, and 0.20\% belongs to ET. Thus, there are one majority class and three minority classes. Since the distribution of class labels in the dataset is highly disproportionate, the dataset suffers from the class imbalance problem. To deal with class imbalance, we used a weighted Dice loss function by taking inspiration from the various Dice loss functions available in the literature \cite{Milletari2016, Sudre2017}. Let ${{g}_{ci}}$ be the ground truth for the $i$-th voxel belonging to the class $c$ and ${{p}_{ci}}$ be the corresponding prediction, $N$ be the total number of voxels, and $L$ be the total number of classes. Then, we define weighted Dice loss function as
\begin{equation}
\mathbf {L}_{dice} = 
1 - \frac{2 \sum_{c=1}^{L} w_c \sum_{i=1}^{N} g_{ci}p_{ci}}
{\sum_{c=1}^{L} w_c (\sum_{i=1}^{N} g_{ci}^2 + \sum_{i=1}^{N} p_{ci}^2)},
\label{eq1}
\end{equation}
where $w_c=1/(\sum_{i=1}^{N} g_{ci})$ is the weight allocated to the class $c$. Thus, a class containing a smaller number of voxels gets more weight. The weights are important in helping to differentiate the three minority classes better. The weighted Dice loss function is designed to deliver good performance for a multi-class segmentation problem like brain tumour segmentation.

\subsection{Data Augmentation}\label{subsec2.5}

The shape, size, and location of brain tumours have massive variations. It can lead to overfitting if we have limited training data. To reduce overfitting and make the network robust to such variability, we enhanced the diversity of the training dataset by using data augmentation. The techniques we used for data augmentation are random rotations, random horizontal flip, and random vertical flip. We have performed these augmentations on the fly, before each epoch, so that the storage requirements does not increase excessively.

\subsection{Incorporating 3D Contextual Information}\label{subsec2.6}

In segmentation, a voxel's class label is highly correlated with that of its neighbouring voxels. Since our network is 2D, it can only capture the correlation of a voxel to its neighbours that are on the same 2D slice. But to capture the correlation of a voxel to its neighbours that are on a different 2D slice we need to incorporate 3D contextual information from axial, sagittal, and coronal views. It can help to determine the class label of a voxel more accurately, thereby improving the performance of our model. We have used two different approaches for combining the 3D contextual information from multiple views. In the first approach, we extracted training patches from axial, sagittal, and coronal views. Then, we used all these extracted patches as input for training a single network. In the second approach, we trained our network in axial, sagittal, and coronal views, respectively. Then we used an ensemble of these three networks for making prediction by taking the average of the softmax outputs of the three networks.

\section{Experiments and Results}\label{sec3}

\subsection{Dataset}\label{subsec3.1}

We carried out our experiments by taking the aid of the BraTS 2018 \cite{Menze2015, Bakas2017, Bakas2017a, Bakas2017b, Bakas2018} dataset. The BraTS training dataset includes pre-operative MRI brain images of 285 patients out of which 210 are for HGG, and 75 are for LGG. The ground truths are also given for the training dataset. The BraTS validation dataset contains MRI images of 66 patients. The ground truths are not given for the validation set, and the online evaluation tool \cite{PennImaging} needs to be used to determine the algorithm's performance on the validation set. For all patients, T1, T1ce, T2, and FLAIR MRI modalities were given. The four MRI modalities for the same patient had been co-registered, and every image had been resampled to 1 mm\textsuperscript{3} isotropic resolution and had been skull stripped. Every MRI image has dimensions of $155\times 240\times 240$. The voxels in the MRI images are grouped into four classes. The NCR and the NET are assigned class label 1, ED is assigned label 2, ET is assigned label 4, and everything else is assigned label 0. Label 3 is not used.

\begin{table*}[t]
\centering
\caption{Quantitative results on the validation set of BraTS 2018.}
\label{tab2}
\renewcommand{\arraystretch}{1.3}
\begin{tabular}{m{5cm}<{\raggedright\arraybackslash}m{1.5cm}<{\centering}m{1.5cm}<{\centering}m{1.5cm}<{\centering}m{1.5cm}<{\centering}m{1.5cm}<{\centering}m{1.5cm}<{\centering}}
\toprule
\multicolumn{1}{c}{\multirow{2}{*}{Method}} & \multicolumn{3}{c}{Dice score} & \multicolumn{3}{c}{Hausdorff distance (95\% quantile)} \\ \cmidrule(lr){2-4} \cmidrule(l){5-7}
\multicolumn{1}{c}{} & ET & WT & TC & ET & WT & TC \\ \midrule
Baseline & 0.76 & 0.89 & 0.81 & 7.29 & 10.39 & 11.60 \\
Baseline (depth of `{U}' shape is 4) & 0.74 & 0.87 & 0.79 & 12.65 & 27.91 & 27.31 \\
Baseline (training patches extracted from  three orthogonal views) & 0.78 & 0.89 & 0.81 & 8.18 & 13.76 & 7.22 \\
Baseline (sagittal view) & 0.75 & 0.88 & 0.80 & 5.23 & 18.31 & 10.69 \\
Baseline (coronal view) & 0.77 & 0.88 & 0.79 & 14.30 & 15.62 & 16.08 \\
Ensemble of three models & 0.79 & 0.90 & 0.82 & 2.99 & 6.28 & 5.90 \\ \bottomrule
\end{tabular}
\end{table*}

\begin{table*}[t]
\centering
\caption{Quantitative results on the dataset of BraTS 2018.}
\label{tab3}
\renewcommand{\arraystretch}{1.3}
\begin{tabular}{m{1.5cm}<{\raggedright\arraybackslash}m{1.5cm}<{\centering}m{1.5cm}<{\centering}m{1.5cm}<{\centering}m{1.5cm}<{\centering}m{1.5cm}<{\centering}m{1.5cm}<{\centering}}
\toprule
\multicolumn{1}{c}{\multirow{2}{*}{Dataset}} & \multicolumn{3}{c}{Sensitivity} & \multicolumn{3}{c}{Specificity} \\ \cmidrule(lr){2-4} \cmidrule(l){5-7}
\multicolumn{1}{c}{} & ET & WT & TC & ET & WT & TC \\ \midrule
Training & 0.86 & 0.93 & 0.91 & 0.999 & 0.997 & 0.998 \\
Validation & 0.81 & 0.89 & 0.79 & 0.997 & 0.995 & 0.998 \\
\bottomrule
\end{tabular}
\end{table*}

\begin{figure*}[b]
\centering{\includegraphics[scale=1]{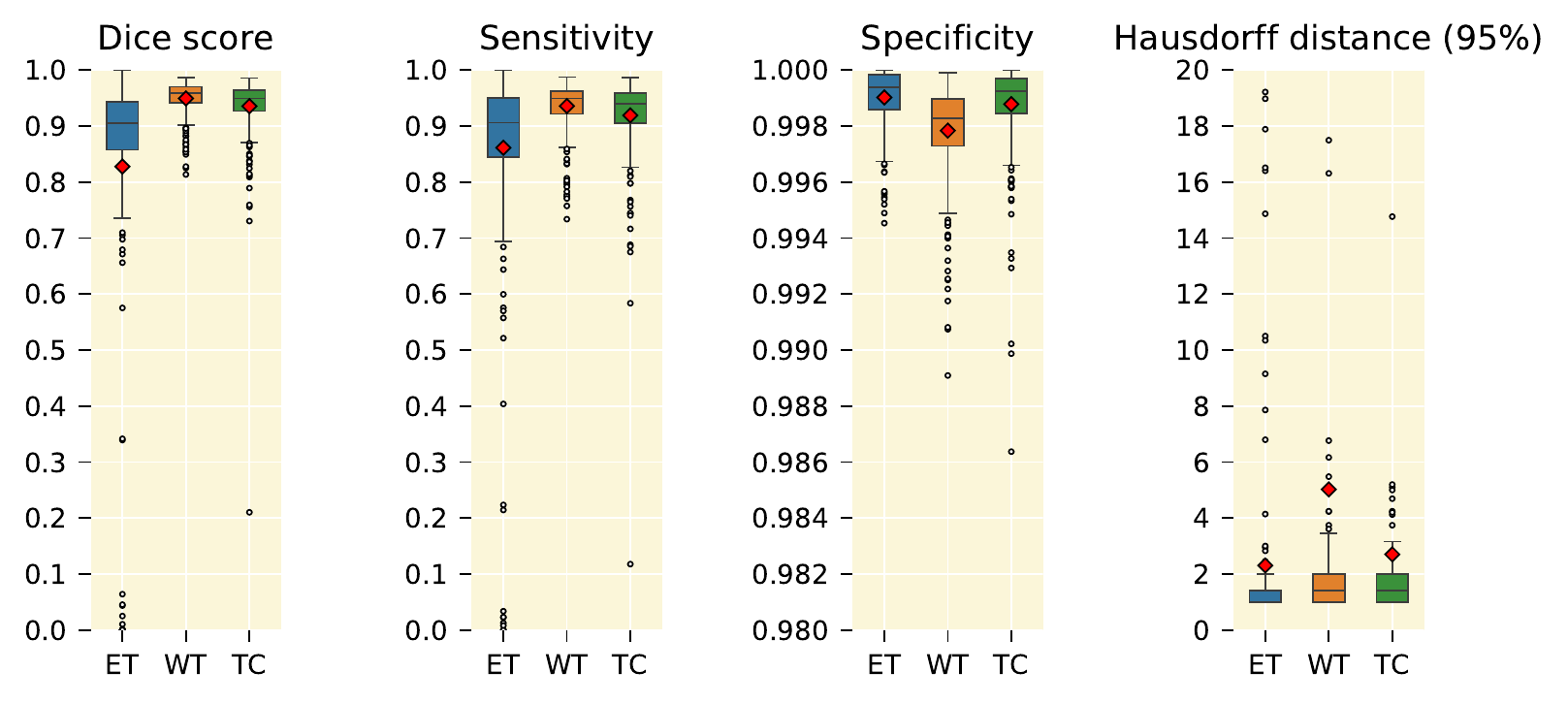}}
\caption{Boxplots for each of the three tumour sub-regions on the BraTS training set. The first column represents the boxplots of Dice score, second column represents the boxplots of sensitivity, third column represents the boxplots of specificity, and fourth column represents the boxplots of 95\% quantile of the Hausdorff distance. The mean is denoted using a red coloured diamond.}
\label{fig2}
\end{figure*}

\begin{figure*}[t]
\centering{\includegraphics[scale=1]{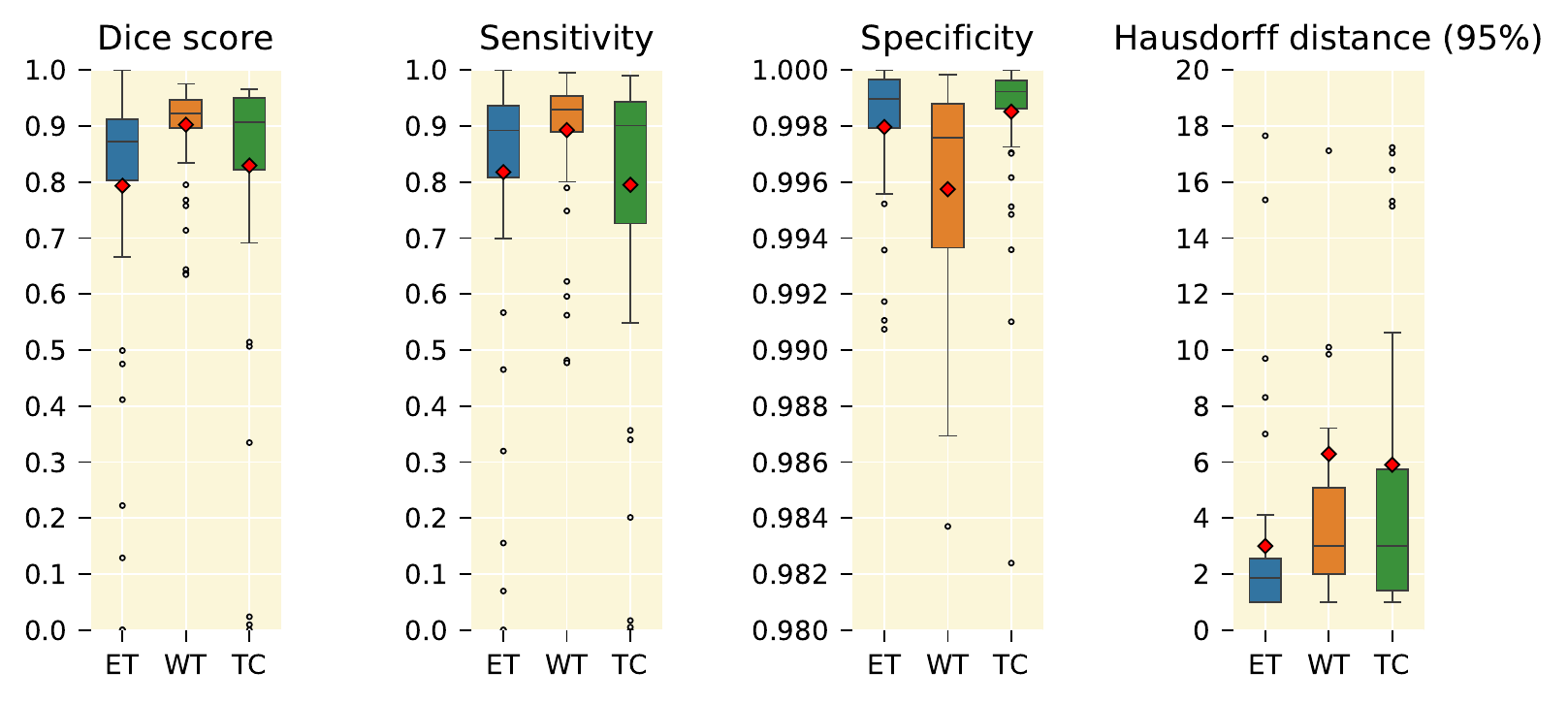}}
\caption{Boxplots for each of the three tumour sub-regions on the BraTS validation set. The first column represents the boxplots of Dice score, second column represents the boxplots of sensitivity, third column represents the boxplots of specificity, and fourth column represents the boxplots of 95\% quantile of the Hausdorff distance. The mean is denoted using a red coloured diamond.}
\label{fig3}
\end{figure*}

\begin{figure*}[t]
\centering{\includegraphics[scale=0.85]{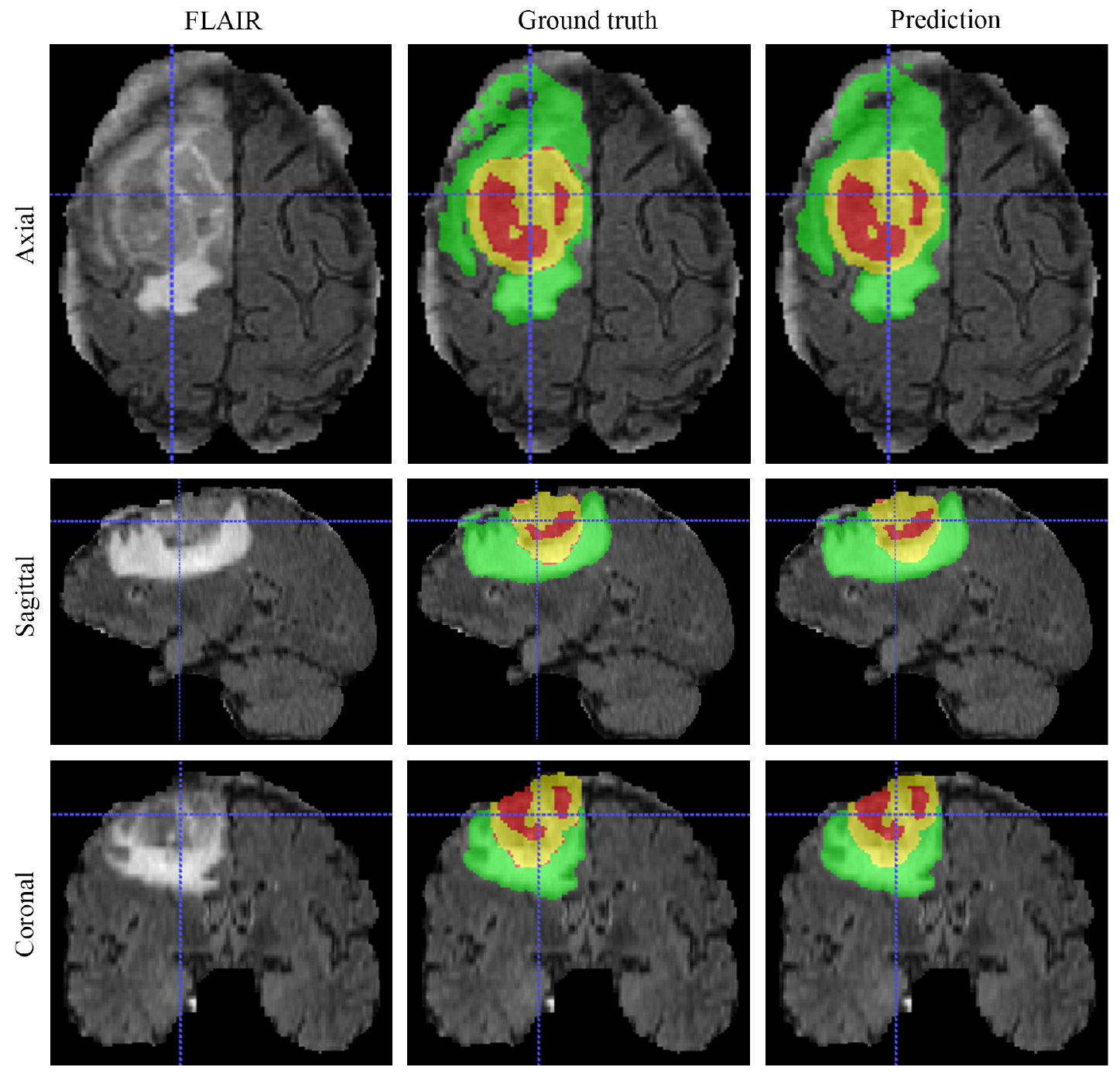}}
\caption{Qualitative result for a patient from the BraTS training set. The left column displays the FLAIR image, the middle column displays the ground truth overlaid over the FLAIR image, and right column displays our predicted segmentation mask overlaid over the FLAIR image. The top row displays the axial slices, middle row displays the sagittal slices, and bottom row displays the coronal slices. Yellow colour represents the ET, red colour represents the NCR and the NET, and green colour represents the ED.}
\label{fig4}
\end{figure*}

\subsection{Results}\label{subsec3.2}

We have used the online evaluation tool \cite{PennImaging} for measuring our method's performance. The online evaluation tool reports the performance of the algorithm with regard to Dice score, sensitivity, specificity, and Hausdorff distance (95\% quantile) for the ET, WT, and TC tumour sub-regions. Quantitative segmentation results with regard to Dice coefficient and 95\% quantile of the Hausdorff distance on the BraTS training set and the BraTS validation set are presented in Table~\ref{tab1} and Table~\ref{tab2}, respectively. In the tables, we use the term baseline to denote our network that used a weighted Dice loss function and was trained in axial view. Next, we increased the depth of the `{U}' shape from 3 to 4 by adding one extra level, and it resulted in a decrease in performance. It happens because increasing the depth of the `{U}' shape beyond certain level results in a massive reduction in image resolution leading to a significant loss of spatial information. Thus, we must always use an optimum depth of the `{U}' shape for achieving excellent performance. We then trained a single model by extracting patches from three orthogonal views. It resulted in little improvement in Dice score for the enhancing tumour but the Hausdorff distance increased for all the three tumour sub-regions. It shows that a single 2D network cannot capture 3D contextual information from multiple views. We also trained our baseline in sagittal and coronal views. Then we used an ensemble of three models trained in the three orthogonal views. It increased the Dice score but more importantly, it helped to bring down the Hausdorff distance. It shows that 3D contextual information can be effectively captured by our ensemble of 2D networks trained in multiple views. We show the quantitative segmentation results on the dataset of BraTS with regard to sensitivity and specificity in Table~\ref{tab3}. In Fig.~\ref{fig2} and Fig.~\ref{fig3} we present the boxplots of the four evaluation metrics for each of three tumour sub-regions on the BraTS training set and BraTS validation set respectively. In the boxplots for Dice score and sensitivity, it is found that the WT sub-region has the lowest spread and the highest mean whereas the ET sub-region has the lowest mean. The boxplots for specificity have the highest spread and lowest mean for the WT sub-region. In the boxplots for the 95\% quantile of Hausdorff distance, the ET sub-region has the lowest spread and the lowest mean.

We show the qualitative result for a patient from the BraTS training set in Fig.~\ref{fig4}. In that figure, yellow colour represents the ET, red colour represents the NCR and the NET, and green colour represents the ET. It is observed in the figure that our model effectively segments the different tumoural structures. We used the ITK-SNAP \cite{Yushkevich2006a, ITK-SNAP} tool for visualising the MRI image.

In Table~\ref{tab4}, we compare the results obtained by our model with those of the state-of-the-art methods on the validation set of BraTS 2018. In comparison to the state-of-the-art techniques that used 3D networks, our approach using 2D network has very less time and memory consumption, and is much simpler and easier to implement; yet it manages to achieve performance that is similar to those of the latest techniques.

\begin{table*}[t]
\centering
\caption{Results obtained by our technique on the validation set of BraTS 2018 (submission id IML) \cite{BraTS18Leaderboard}. We also display the results of the state-of-the-art methods for comparison.}
\label{tab4}
\renewcommand{\arraystretch}{1.3}
\begin{tabular}{m{2.75cm}<{\raggedright\arraybackslash}m{1.5cm}<{\centering}m{1.5cm}<{\centering}m{1.5cm}<{\centering}m{1.5cm}<{\centering}m{1.5cm}<{\centering}m{1.5cm}<{\centering}}
\toprule
\multicolumn{1}{c}{\multirow{2}{*}{Methods}} & \multicolumn{3}{c}{Dice score} & \multicolumn{3}{c}{Hausdorff distance (95\% quantile)} \\ \cmidrule(lr){2-4} \cmidrule(l){5-7}
\multicolumn{1}{c}{} & ET & WT & TC & ET & WT & TC \\ \midrule
Proposed & 0.79 & 0.90 & 0.82 & 2.99 & 6.28 & 5.90 \\
Myronenko \cite{Myronenko2019} & 0.82 & 0.91 & 0.86 & 3.92 & 4.51 & 6.85 \\
Isensee et al. \cite{Isensee2019} & 0.80 & 0.91 & 0.86 & 2.41 & 4.27 & 6.52 \\
McKinley et al. \cite{McKinley2019} & 0.79 & 0.90 & 0.84 & 3.55 & 4.17 & 4.93 \\
Zhou et al. \cite{Zhou2019} & 0.81 & 0.90 & 0.86 & 2.71 & 4.17 & 6.54 \\
\bottomrule
\end{tabular}
\end{table*}

\section{Conclusions}\label{sec4}

In this work, we have described our automated glioma segmentation method using 2D CNNs that are based on U-Net \cite{Ronneberger2015}. To effectively deal with class imbalance problem, we have used a weighted Dice loss function. We investigated and found that increasing the depth of the `{U}' shape beyond certain level results in a massive reduction in image resolution leading to loss of spatial information and a decrease in performance. Thus, for CNNs based on U-Net, we must always use an optimum depth of the `{U}' shape for achieving the best results possible. We investigated the effectiveness of two different approaches for incorporating 3D contextual information in a 2D network to enhance its performance. In the first approach, we used a single model that is trained by extracting patches from three orthogonal views but found that this approach resulted in a decrease in performance. In the second approach, we used an ensemble of three models trained in multiple views, and this approach resulted in improvement in performance. We found that 3D contextual information cannot be captured by a single 2D network that is trained with patches extracted from multiple views whereas an ensemble of three 2D networks trained in multiple views can effectively capture the information and deliver much better performance. Finally, we obtained Dice scores of 0.79, 0.90 and 0.82 for the ET, WT, and TC, respectively on the BraTS 2018 validation set and its results are similar to those of the latest techniques. Our method using 2D network consumes very less time and memory, and is much simpler and easier to implement compared to the most recent techniques that used 3D networks; still, it manages to achieve comparable performance to the latest methods because of the incorporation of 3D contextual information from multiple views. It shows that a 2D network that has incorporated 3D contextual information from orthogonal views can deliver comparable performance to a 3D network.



\begin{thebibliography}{99}

\bibitem{DeAngelis2001}
DeAngelis, L.M.: `{Brain Tumors}', \emph{New England Journal of Medicine},
  2001, \textbf{344}, (2), pp.~114--123.
\newblock Available from:
  \url{http://www.nejm.org/doi/abs/10.1056/NEJM200101113440207}

\bibitem{Ohgaki2005}
Ohgaki, H., Kleihues, P.: `{Population-Based Studies on Incidence, Survival
  Rates, and Genetic Alterations in Astrocytic and Oligodendroglial Gliomas}',
  \emph{Journal of Neuropathology {\&} Experimental Neurology},  2005,
  \textbf{64}, (6), pp.~479--489.
\newblock Available from:
  \url{https://academic.oup.com/jnen/article-lookup/doi/10.1093/jnen/64.6.479}

\bibitem{Goodenberger2012}
Goodenberger, M.L., Jenkins, R.B.: `{Genetics of adult glioma}', \emph{Cancer
  genetics},  2012, \textbf{205}, (12), pp.~613--21.
\newblock Available from: \url{http://www.ncbi.nlm.nih.gov/pubmed/23238284}

\bibitem{Bauer2013}
Bauer, S., Wiest, R., Nolte, L.P., Reyes, M.: `{A survey of MRI-based medical
  image analysis for brain tumor studies}', \emph{Physics in Medicine and
  Biology},  2013, \textbf{58}, (13), pp.~R97--R129.
\newblock Available from:
  \url{http://stacks.iop.org/0031-9155/58/i=13/a=R97?key=crossref.f5d87890e52190a8b3681035112ecdfa}

\bibitem{Louis2016}
Louis, D.N., Perry, A., Reifenberger, G., von Deimling, A., Figarella.Branger,
  D., Cavenee, W.K., et~al.: `{The 2016 World Health Organization
  Classification of Tumors of the Central Nervous System: a summary}',
  \emph{Acta Neuropathologica},  2016, \textbf{131}, (6), pp.~803--820.
\newblock Available from:
  \url{http://link.springer.com/10.1007/s00401-016-1545-1}

\bibitem{Menze2015}
Menze, B.H., Jakab, A., Bauer, S., Kalpathy.Cramer, J., Farahani, K., Kirby,
  J., et~al.: `{The Multimodal Brain Tumor Image Segmentation Benchmark
  (BRATS)}', \emph{IEEE Transactions on Medical Imaging},  2015, \textbf{34},
  (10), pp.~1993--2024.
\newblock Available from: \url{http://ieeexplore.ieee.org/document/6975210/}

\bibitem{Isin2016}
I\c s\i n, A., Direko\u glu, C., \c Sah, M.: `{Review of MRI-based Brain Tumor Image
  Segmentation Using Deep Learning Methods}', \emph{Procedia Computer Science},
   2016, \textbf{102}, pp.~317--324.
\newblock Available from:
  \url{https://www.sciencedirect.com/science/article/pii/S187705091632587X}

\bibitem{Bakas2018}
Bakas, S., Reyes, M., Jakab, A., Bauer, S., Rempfler, M., Crimi, A., et~al.:
  `{Identifying the Best Machine Learning Algorithms for Brain Tumor
  Segmentation, Progression Assessment, and Overall Survival Prediction in the
  BRATS Challenge}', \emph{arXiv:181102629 [cs.CV]},  2018. Available from:
  \url{http://arxiv.org/abs/1811.02629}

\bibitem{Menze2010}
Menze, B.H., van Leemput, K., Lashkari, D., Weber, M.A., Ayache, N., Golland,
  P.
\newblock `{A Generative Model for Brain Tumor Segmentation in Multi-Modal
  Images}'.
\newblock In: Jiang, T., Navab, N., Pluim, J.P.W., Viergever, M.A., editors.
  Medical Image Computing and Computer-Assisted Intervention - MICCAI 2010.
  Springer, Berlin, Heidelberg,  2010. pp.~ 151--159.
\newblock Available from:
  \url{http://link.springer.com/10.1007/978-3-642-15745-5_19}

\bibitem{Pereira2016}
Pereira, S., Pinto, A., Alves, V., Silva, C.A.: `{Brain Tumor Segmentation
  Using Convolutional Neural Networks in MRI Images}', \emph{IEEE Transactions
  on Medical Imaging},  2016, \textbf{35}, (5), pp.~1240--1251.
\newblock Available from: \url{http://ieeexplore.ieee.org/document/7426413/}

\bibitem{Kamnitsas2017a}
Kamnitsas, K., Ledig, C., Newcombe, V.F.J., Simpson, J.P., Kane, A.D., Menon,
  D.K., et~al.: `{Efficient multi-scale 3D CNN with fully connected CRF for
  accurate brain lesion segmentation}', \emph{Medical Image Analysis},  2017,
  \textbf{36}, pp.~61--78.
\newblock Available from:
  \url{https://www.sciencedirect.com/science/article/pii/S1361841516301839}

\bibitem{Kamnitsas2018}
Kamnitsas, K., Bai, W., Ferrante, E., McDonagh, S., Sinclair, M., Pawlowski,
  N., et~al.
\newblock `{Ensembles of Multiple Models and Architectures for Robust Brain
  Tumour Segmentation}'.
\newblock In: Crimi, A., Bakas, S., Kuijf, H., Menze, B., Reyes, M., editors.
  Brainlesion: Glioma, Multiple Sclerosis, Stroke and Traumatic Brain Injuries.
  Springer, Cham,  2018. pp.~ 450--462.
\newblock Available from:
  \url{http://link.springer.com/10.1007/978-3-319-75238-9_38}

\bibitem{Kamnitsas2015}
Kamnitsas, K., Chen, L., Ledig, C., Rueckert, D., Glocker, B.
\newblock `{Multi-Scale 3D Convolutional Neural Networks for Lesion
  Segmentation in Brain MRI}'.
\newblock In: Proceedings of ISLES-MICCAI. (,  2015. Available from:
  \url{http://www.isles-challenge.org/ISLES2015/articles/kamnk1.pdf}

\bibitem{Long2015}
Long, J., Shelhamer, E., Darrell, T.
\newblock `{Fully convolutional networks for semantic segmentation}'.
\newblock In: 2015 IEEE Conference on Computer Vision and Pattern Recognition
  (CVPR). (IEEE,  2015. pp.~ 3431--3440.
\newblock Available from: \url{http://ieeexplore.ieee.org/document/7298965/}

\bibitem{Ronneberger2015}
Ronneberger, O., Fischer, P., Brox, T.
\newblock `{U-Net: Convolutional Networks for Biomedical Image Segmentation}'.
\newblock In: Navab, N., Hornegger, J., Wells, W.M., Frangi, A.F., editors.
  Medical Image Computing and Computer-Assisted Intervention - MICCAI 2015.
  Springer, Cham,  2015. pp.~ 234--241.
\newblock Available from:
  \url{http://link.springer.com/10.1007/978-3-319-24574-4_28}

\bibitem{Wang2018d}
Wang, G., Li, W., Ourselin, S., Vercauteren, T.
\newblock `{Automatic Brain Tumor Segmentation Using Cascaded Anisotropic
  Convolutional Neural Networks}'.
\newblock In: Crimi, A., Bakas, S., Kuijf, H., Menze, B., Reyes, M., editors.
  Brainlesion: Glioma, Multiple Sclerosis, Stroke and Traumatic Brain Injuries.
  Springer, Cham,  2018. pp.~ 178--190.
\newblock Available from:
  \url{http://link.springer.com/10.1007/978-3-319-75238-9_16}

\bibitem{Isensee2018}
Isensee, F., Kickingereder, P., Wick, W., Bendszus, M., Maier.Hein, K.H.
\newblock `{Brain Tumor Segmentation and Radiomics Survival Prediction:
  Contribution to the BRATS 2017 Challenge}'.
\newblock In: Crimi, A., Bakas, S., Kuijf, H., Menze, B., Reyes, M., editors.
  Brainlesion: Glioma, Multiple Sclerosis, Stroke and Traumatic Brain Injuries.
  Springer, Cham,  2018. pp.~ 287--297.
\newblock Available from:
  \url{http://link.springer.com/10.1007/978-3-319-75238-9_25}

\bibitem{He2016a}
He, K., Zhang, X., Ren, S., Sun, J.
\newblock `{Identity Mappings in Deep Residual Networks}'.
\newblock In: Leibe, B., Matas, J., Sebe, N., Welling, M., editors. Computer
  Vision -- ECCV 2016. Springer, Cham,  2016. pp.~ 630--645.
\newblock Available from:
  \url{http://link.springer.com/10.1007/978-3-319-46493-0_38}

\bibitem{Myronenko2019}
Myronenko, A.
\newblock `{3D MRI Brain Tumor Segmentation Using Autoencoder Regularization}'.
\newblock In: Crimi, A., Bakas, S., Kuijf, H., Keyvan, F., Reyes, M., van
  Walsum, T., editors. Brainlesion: Glioma, Multiple Sclerosis, Stroke and
  Traumatic Brain Injuries. Springer, Cham,  2019. pp.~ 311--320.
\newblock Available from:
  \url{http://link.springer.com/10.1007/978-3-030-11726-9_28}

\bibitem{Isensee2019}
Isensee, F., Kickingereder, P., Wick, W., Bendszus, M., Maier.Hein, K.H.
\newblock `{No New-Net}'.
\newblock In: Crimi, A., Bakas, S., Kuijf, H., Keyvan, F., Reyes, M., van
  Walsum, T., editors. Brainlesion: Glioma, Multiple Sclerosis, Stroke and
  Traumatic Brain Injuries. Springer, Cham,  2019. pp.~ 234--244.
\newblock Available from:
  \url{http://link.springer.com/10.1007/978-3-030-11726-9_21}

\bibitem{Cicek2016}
{\c{C}}i{\c{c}}ek, {\"{O}}., Abdulkadir, A., Lienkamp, S.S., Brox, T.,
  Ronneberger, O.
\newblock `{3D U-Net: Learning Dense Volumetric Segmentation from Sparse
  Annotation}'.
\newblock In: Ourselin, S., Joskowicz, L., Sabuncu, M.R., Unal, G., Wells, W.,
  editors. Medical Image Computing and Computer-Assisted Intervention - MICCAI
  2016. Springer, Cham,  2016. pp.~ 424--432.
\newblock Available from:
  \url{http://link.springer.com/10.1007/978-3-319-46723-8_49}

\bibitem{McKinley2019}
McKinley, R., Meier, R., Wiest, R.
\newblock `{Ensembles of Densely-Connected CNNs with Label-Uncertainty for
  Brain Tumor Segmentation}'.
\newblock In: Crimi, A., Bakas, S., Kuijf, H., Keyvan, F., Reyes, M., van
  Walsum, T., editors. Brainlesion: Glioma, Multiple Sclerosis, Stroke and
  Traumatic Brain Injuries. Springer, Cham,  2019. pp.~ 456--465.
\newblock Available from:
  \url{http://link.springer.com/10.1007/978-3-030-11726-9_40}

\bibitem{Huang2017}
Huang, G., Liu, Z., van~der Maaten, L., Weinberger, K.Q.
\newblock `{Densely Connected Convolutional Networks}'.
\newblock In: 2017 IEEE Conference on Computer Vision and Pattern Recognition
  (CVPR). (IEEE,  2017. pp.~ 2261--2269.
\newblock Available from: \url{http://ieeexplore.ieee.org/document/8099726/}

\bibitem{Zhou2019}
Zhou, C., Chen, S., Ding, C., Tao, D.
\newblock `{Learning Contextual and Attentive Information for Brain Tumor
  Segmentation}'.
\newblock In: Crimi, A., Bakas, S., Kuijf, H., Keyvan, F., Reyes, M., van
  Walsum, T., editors. Brainlesion: Glioma, Multiple Sclerosis, Stroke and
  Traumatic Brain Injuries. Springer, Cham,  2019. pp.~ 497--507.
\newblock Available from:
  \url{http://link.springer.com/10.1007/978-3-030-11726-9_44}

\bibitem{Milletari2016}
Milletari, F., Navab, N., Ahmadi, S.A.
\newblock `{V-Net: Fully Convolutional Neural Networks for Volumetric Medical
  Image Segmentation}'.
\newblock In: 2016 Fourth International Conference on 3D Vision (3DV). (IEEE,
  2016. pp.~ 565--571.
\newblock Available from: \url{http://ieeexplore.ieee.org/document/7785132/}

\bibitem{Sudre2017}
Sudre, C.H., Li, W., Vercauteren, T., Ourselin, S., {Jorge Cardoso}, M.
\newblock `{Generalised Dice Overlap as a Deep Learning Loss Function for
  Highly Unbalanced Segmentations}'.
\newblock In: Cardoso, M.J., Arbel, T., Carneiro, G., Syeda.Mahmood, T.,
  Tavares, J.M.R.S., Moradi, M., et~al., editors. Deep Learning in Medical
  Image Analysis and Multimodal Learning for Clinical Decision Support.
  Springer, Cham,  2017. pp.~ 240--248.
\newblock Available from:
  \url{http://link.springer.com/10.1007/978-3-319-67558-9_28}

\bibitem{Ioffe2015}
Ioffe, S., Szegedy, C.
\newblock `{Batch Normalization: Accelerating Deep Network Training by Reducing
  Internal Covariate Shift}'.
\newblock In: Bach, F., Blei, D., editors. Proceedings of the 32nd
  International Conference on Machine Learning. vol.~37 of \emph{Proceedings of
  Machine Learning Research}. (Lille, France: PMLR,  2015. pp.~ 448--456.
\newblock Available from: \url{http://proceedings.mlr.press/v37/ioffe15.html}

\bibitem{Keras}
`{Home - Keras Documentation}'.
\newblock Available from: \url{https://keras.io/}, accessed July 2019

\bibitem{Abadi2016}
Abadi, M., Barham, P., Chen, J., Chen, Z., Davis, A., Dean, J., et~al.
\newblock `{TensorFlow: A system for large-scale machine learning}'.
\newblock In: OSDI'16 Proceedings of the 12th USENIX conference on Operating
  Systems Design and Implementation. 2016. pp.~ 265--283.
\newblock Available from:
  \url{https://www.usenix.org/conference/osdi16/technical-sessions/presentation/abadi}

\bibitem{Kingma2015}
Kingma, D.P., Ba, J.
\newblock `{Adam: A Method for Stochastic Optimization}'.
\newblock In: Proceedings of the 3rd International Conference on Learning
  Representations (ICLR), 2015. Available from:
  \url{http://arxiv.org/abs/1412.6980}

\bibitem{Bakas2017}
Bakas, S., Akbari, H., Sotiras, A., Bilello, M., Rozycki, M., Kirby, J.S.,
  et~al.: `{Advancing The Cancer Genome Atlas glioma MRI collections with
  expert segmentation labels and radiomic features}', \emph{Scientific Data},
  2017, \textbf{4}, pp.~1--13.
\newblock Available from: \url{http://www.nature.com/articles/sdata2017117}

\bibitem{Bakas2017a}
Bakas, S., Akbari, H., Sotiras, A., Bilello, M., Rozycki, M., Kirby, J.,
  et~al.. `{Segmentation Labels and Radiomic Features for the Pre-operative
  Scans of the TCGA-GBM collection}', 2017.
\newblock Available from:
  \url{https://wiki.cancerimagingarchive.net/display/DOI/Segmentation+Labels+and+Radiomic+Features+for+the+Pre-operative+Scans+of+the+TCGA-GBM+collection}

\bibitem{Bakas2017b}
Bakas, S., Akbari, H., Sotiras, A., Bilello, M., Rozycki, M., Kirby, J.,
  et~al.. `{Segmentation Labels and Radiomic Features for the Pre-operative
  Scans of the TCGA-LGG collection}', 2017.
\newblock Available from:
  \url{https://wiki.cancerimagingarchive.net/display/DOI/Segmentation+Labels+and+Radiomic+Features+for+the+Pre-operative+Scans+of+the+TCGA-LGG+collection}

\bibitem{PennImaging}
`{Penn Imaging -- Home}'.
\newblock Available from: \url{https://ipp.cbica.upenn.edu/}, accessed July 2019

\bibitem{Yushkevich2006a}
Yushkevich, P.A., Piven, J., Hazlett, H.C., Smith, R.G., Ho, S., Gee, J.C.,
  et~al.: `{User-guided 3D active contour segmentation of anatomical
  structures: Significantly improved efficiency and reliability}',
  \emph{NeuroImage},  2006, \textbf{31}, (3), pp.~1116--1128.
\newblock Available from:
  \url{https://www.sciencedirect.com/science/article/pii/S1053811906000632}

\bibitem{ITK-SNAP}
`{ITK-SNAP Home}'.
\newblock Available from: \url{http://www.itksnap.org/pmwiki/pmwiki.php}, accessed July 2019

\bibitem{BraTS18Leaderboard}
`{MICCAI-BraTS 2018 Leaderboard}'.
\newblock Available from:
  \url{https://www.cbica.upenn.edu/BraTS18/lboardValidation.html}, accessed July 2019

\end{thebibliography}


\end{document}